\documentclass[12pt,a4paper]{article}
\usepackage[utf8]{inputenc}
\usepackage[T1]{fontenc}
\usepackage{graphicx}	
\usepackage{amsmath}	
\usepackage{amssymb}
\usepackage{multirow}
\usepackage{amsfonts}
\usepackage{mathtools}
\usepackage{hyperref}
\usepackage{authblk}

\title{Air Resistance From the Acceleration of a Falling Smartphone}

\author[1]{Fernando Saliby}
\affil[1]{Departamento de Ci\^encias da Natureza, Universidade Federal Fluminense, Rua Recife, Lotes, 1-7, Jardim Bela Vista, Rio das Ostras - RJ, 28895-532, Brasil}
\affil[ ]{\texttt{fsimoni@id.uff.br}}

\begin{document}

\maketitle

\begin{abstract}
This study investigates the motion of a falling smartphone under the influence of air drag using acceleration data collected by its built-in accelerometer. The proper acceleration profiles demonstrate the suitability of the turbulent drag model in capturing the motion dynamics during both upward and downward phases. This approach provides an effective and accessible method for exploring fluid dynamics concepts in an educational context.
\end{abstract}

\section{Introduction}\label{sec:intro}

The study of falling objects under the influence of air resistance is a common topic in introductory experimental physics courses~\cite{pagonis97,mohazzabi2011,mohazzabi2018}, providing valuable insights into the interplay between gravitational forces and fluid dynamics. However, accurately modeling air drag presents challenges due to the numerous variables involved, such as the object's shape, size, orientation, velocity, and the properties of the surrounding fluid~\cite{kundu2010fluid}. These factors make air resistance a nuanced subject with both theoretical and practical relevance.

The intricate nature of air-object interactions gave rise to the specialized field of aerodynamics~\cite{anderson2010}. This area often relies on advanced tools such as wind tunnels and computational fluid dynamics to tackle its demanding problems. Nevertheless, in basic physics courses, air drag is addressed through simplified models that capture the essence of practical scenarios. Among these, the turbulent flow model stands out as particularly relevant for objects of moderate size moving at velocities where Reynolds numbers exceed critical values, such as falling smartphones~\cite{parker77,pettersson19}.

A straightforward way to explore air drag experimentally is by filming falling objects and analyzing their position-time data using video analysis software~\cite{Wijaya_2019}. Although this method is intuitive and accessible, it has limitations in capturing detailed dynamic behavior, as instantaneous acceleration changes are difficult to extract from position data. To overcome this limitation, we propose using smartphones' built-in accelerometers to measure proper acceleration during free-fall experiments. This approach highlights the key characteristics of turbulent drag, which are evident in the acceleration profiles.

This method also offers valuable educational opportunities by providing a simple experimental setup---dropping the smartphone---which facilitates the exploration of real data and fundamental physical concepts. The inherent variability in experimental conditions, such as initial velocities, tilt, or potential rotations of the phone, emphasizes the role of experimental uncertainties and the importance of repeated measurements. These features offer a comprehensive view of the experimental process, emphasizing hands-on experimentation, data selection and analysis, and the integration of modern technology.

The remainder of this paper is organized as follows: Section~\ref{sec:model} presents the theoretical model for the smartphone's motion under air resistance, focusing on the turbulent regime. Section~\ref{sec:Data_Acquisition} describes the experimental setup and data collection using the smartphone's accelerometer. In Section~\ref{sec:Data_Analysis}, we outline the methodology and present the results. Section~\ref{sec:Discussions_and_Conclusions} discusses the findings and the effectiveness of the experimental approach.

The analysis code (written in Python\footnote{\url{https://www.python.org/}} using NumPy\footnote{\url{https://numpy.org/}}, Matplotlib\footnote{\url{https://matplotlib.org/}}, and SciPy\footnote{\url{https://scipy.org/}}) and experimental data from this study are available on GitHub\footnote{\url{https://github.com/fsaliby/Air_Resistance_From_the_Acceleration_of_a_Falling_Smartphone}}.

\section{Turbulent Drag Model}\label{sec:model}

As the smartphone moves through the air, it experiences a drag force given by:
\begin{equation}\label{eq:drag_model}
\vec{F}_D = k |\vec{v}_p| \vec{v}_p,
\end{equation}
where \(k\) is a constant that depends on the object's shape, size, orientation, and the fluid's properties. The term \(\vec{v}_p\) denotes the velocity of the air as measured in the smartphone's reference frame. This formulation differs from the conventional drag force equation, which uses the velocity relative to the fluid at rest and includes a negative sign.

To model the air velocity in the smartphone's frame, \( \vec{v}_p \), we apply Newton's second law in an inertial reference frame fixed to the Earth's surface. This is necessary because the accelerating smartphone constitutes a non-inertial frame. We make the following assumptions:

\begin{itemize}
\item The \(z\)-axis points upward in the Earth frame.
\item The smartphone's motion is purely vertical in this frame, i.e., \( \vec{v} = v \hat{z} \).
\item The smartphone's screen faces upward throughout the entire motion, so \( \hat{z} \cdot \hat{z}_{p} = 1 \), where \( \hat{z}_{p} \) denotes the unit vector perpendicular to the smartphone's screen~\cite{vogt12}.
\end{itemize}
Assuming the air is stationary with respect to the Earth, we have \( \vec{v}_p = -\vec{v} \).

With this setup, Newton's second law, in the Earth's reference frame, provides the following smartphone's acceleration:
\begin{equation}
a = -g - \frac{k}{m} |v| v ,
\label{eq:accel_earth}
\end{equation}
where \( g \) is the acceleration due to gravity, and \( m \) is the mass of the smartphone. The second term represents the proper acceleration, \( a_p \), measured by the accelerometer, which comes directly from the drag force (Eq.~\ref{eq:drag_model}):
\begin{equation}
a_p = \frac{F_D}{m} = - \frac{k}{m} |v| v .
\end{equation}
Therefore, we can write:
\begin{equation}
a = -g + a_p .
\label{eq:accel_earth_proper}
\end{equation}
This proper acceleration, \( a_p \), forms the basis of our model and will be compared with the accelerometer measurements.

From Eq.~\eqref{eq:accel_earth}, we can derive an important consequence of air resistance during fall: the existence of a finite terminal velocity. As an object falls, the drag force increases with velocity. Eventually, the drag force equals the gravitational force, resulting in zero net acceleration and a constant velocity. This equilibrium velocity is known as the terminal velocity and is given by:
\begin{equation}\label{eq:terminal_velocity}
|v(t \to \infty)| v(t \to \infty) = -\frac{mg}{k}.
\end{equation}
The negative sign indicates that the terminal velocity is directed downward. We define the terminal speed as the magnitude of this velocity:
\begin{equation}\label{eq:terminal_velocity_magnitude}
v_t = \sqrt{\frac{mg}{k}}.
\end{equation}

By substituting the expression for \( v_t \) into the previous equations, we can rewrite the acceleration terms as:
\begin{align} 
a &= -g \left(1 + \frac{|v| v}{v_t^2}\right), \label{eq:accel_earth_vt} \\
a_p &= -g \frac{|v| v}{v_t^2} . 
\label{eq:accel_proper_vt}
\end{align}
These expressions clearly show the relationship between the velocity \( v \) and the terminal speed \( v_t \), highlighting that as the velocity approaches \( -v_t \), the total acceleration \( a \) approaches zero. The proper acceleration \( a_{p} \) reflects only the effects of air drag. As \( v \) approaches \( -v_t \), \( a_p \) approaches \( g \), creating an accelerated frame that, from the perspective of the accelerometer, is equivalent to being at rest in Earth's gravitational field.

The solution of Eq.~\eqref{eq:accel_earth_vt} depends on the initial conditions. For upward motion (\(v_0 > 0\)), we have:
\begin{equation}\label{eq:v_turbulent_upward}
v(t) = v_t \tan\left( \arctan\left( \frac{v_0}{v_t} \right) - \frac{g t}{v_t} \right).
\end{equation}
During the upward phase, the object decelerates due to both gravity and air resistance until reaching its maximum height. This occurs at the apex, where the velocity becomes zero at time:
\begin{equation}\label{eq:t_apex_turbulent}
t_{\text{apex}} = \frac{v_t}{g} \arctan\left( \frac{v_0}{v_t} \right).
\end{equation}

For downward motion, after the object passes its apex, the velocity is described by:
\begin{equation}\label{eq:v_turbulent_downward}
v(t) = -v_t \tanh\left[ g \left( \frac{t - t_{\text{apex}}}{v_t} \right) \right].
\end{equation}
This equation captures the gradual approach of the velocity to its terminal value (\(-v_t\)) as the object continues to fall under turbulent drag.

For initial downward motion ($v_0 < 0$), the velocity is described by:
\begin{equation}\label{eq:v_turbulent_negative_v0}
v(t) = -v_t \tanh\left[ \text{arctanh}\left( \frac{v_0}{-v_t} \right) + \frac{g t}{v_t} \right].
\end{equation}
These equations complete our theoretical framework for analyzing the acceleration data from both experimental setups.

\subsection{Characteristics of Proper Acceleration}\label{sec:accel_proper_characteristics}

\begin{figure}[t]
    \centering
    \includegraphics[width=0.8\textwidth]{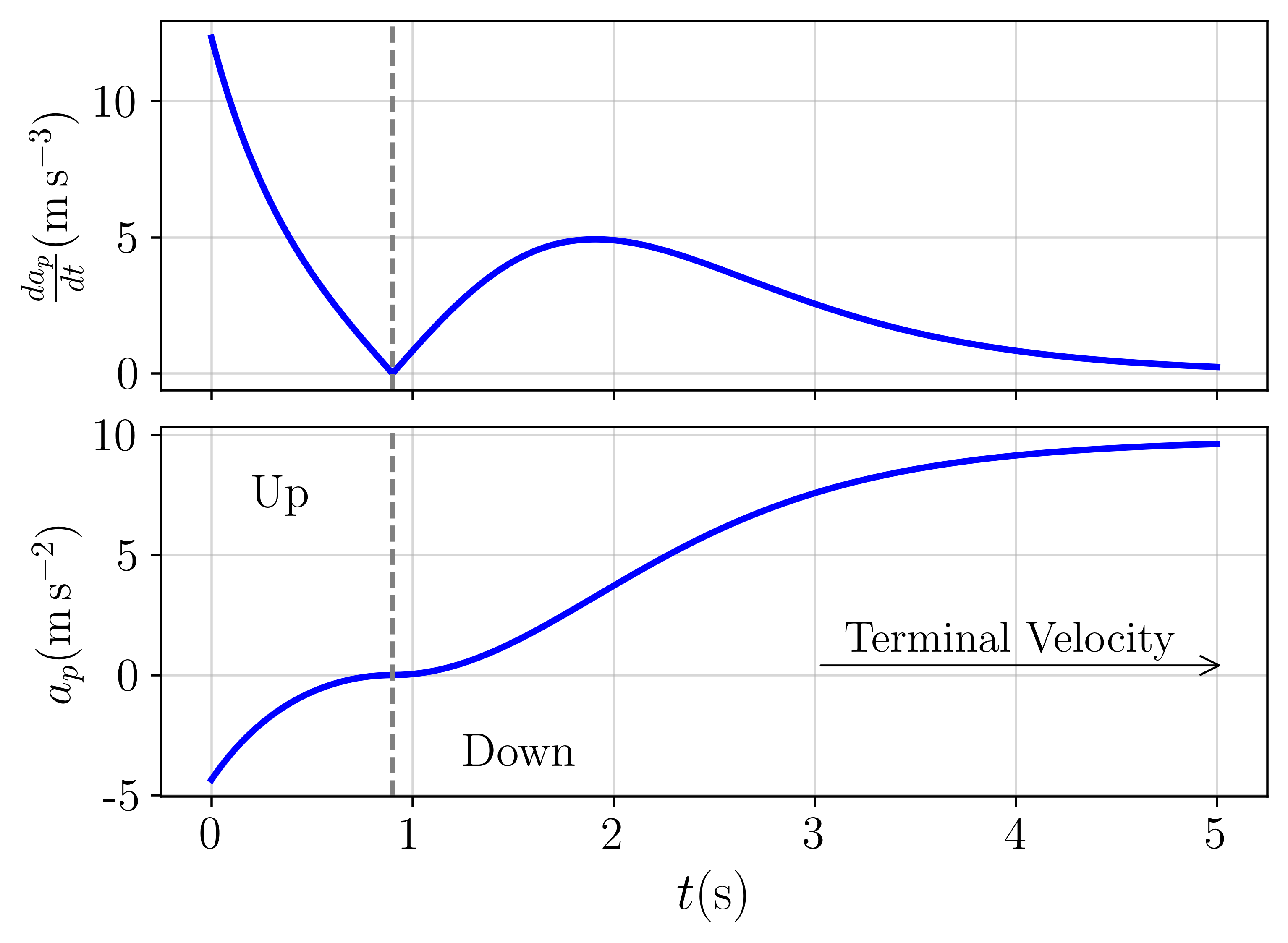}
    \caption{Proper acceleration \( a_p(t) \) (bottom panel) and its derivative \( \frac{da_p}{dt} \) (top panel) for turbulent flow. The gray vertical dashed line indicates the apex time (\(t_{\text{apex}} = 0.90 \, \text{s}\)).}
    \label{fig:ap_and_derivative_n}
\end{figure}

The bottom plot of Fig.~\ref{fig:ap_and_derivative_n} illustrates the proper acceleration \( a_{p} \)---the acceleration measured by the smartphone's accelerometer. The plot uses an initial velocity of \( v_0 = 10 \, \text{m} \, \text{s}^{-1} \) and a terminal velocity of \( v_t = 15 \, \text{m} \, \text{s}^{-1} \). The curve highlights how the proper acceleration evolves over time, showing distinct patterns of concavity and rate of change during different phases of motion.

The proper acceleration curve displays three distinct regions with characteristic behavior:
\begin{itemize}
    \item During the upward motion, the curve shows negative concavity as both gravity and drag force decelerate the object.
    
    \item At the apex, where the velocity becomes zero, the concavity inverts to positive, marking the transition to downward motion.
    
    \item As the object approaches terminal velocity (approximately two seconds into the fall), the curve becomes concave down again, marking the final stage of motion.
\end{itemize}

The top plot of Fig.~\ref{fig:ap_and_derivative_n} shows the derivative of the proper acceleration, also known as jerk~\cite{Eager_2016}, given by:
\begin{equation}\label{eq:a_p_derivative}
\frac{da_p}{dt} = -2 g \frac{|v|}{v_{t}^2} a,
\end{equation}
where \( a \) is the total acceleration of the object in the Earth's frame, accounting for both gravity and drag effects, as described by Eq.~\eqref{eq:accel_earth_vt}. The negative sign ensures that the derivative of the proper acceleration is always positive. As the object approaches terminal velocity, its proper acceleration converges to \( g \), and its derivative approaches zero, indicating that the system is reaching a steady state where the drag force becomes constant.

The second derivative of the proper acceleration, also known as snap~\cite{Eager_2016}, provides further insight into the behavior around the apex. At the apex, where the velocity vanishes, the second derivative is discontinuous and changes sign, indicating an inflection point. Specifically:
\begin{equation}
\left. \frac{d^{2}a_p}{dt^2} \right|_{t_{\text{apex}}} = -2 \frac{g^3}{v_t^2} \, \text{sign}(v),
\end{equation}
where \(\text{sign}(v)\) indicates the direction of the velocity before and after the apex.

\begin{figure}[t]
    \centering
    \includegraphics[width=0.8\textwidth]{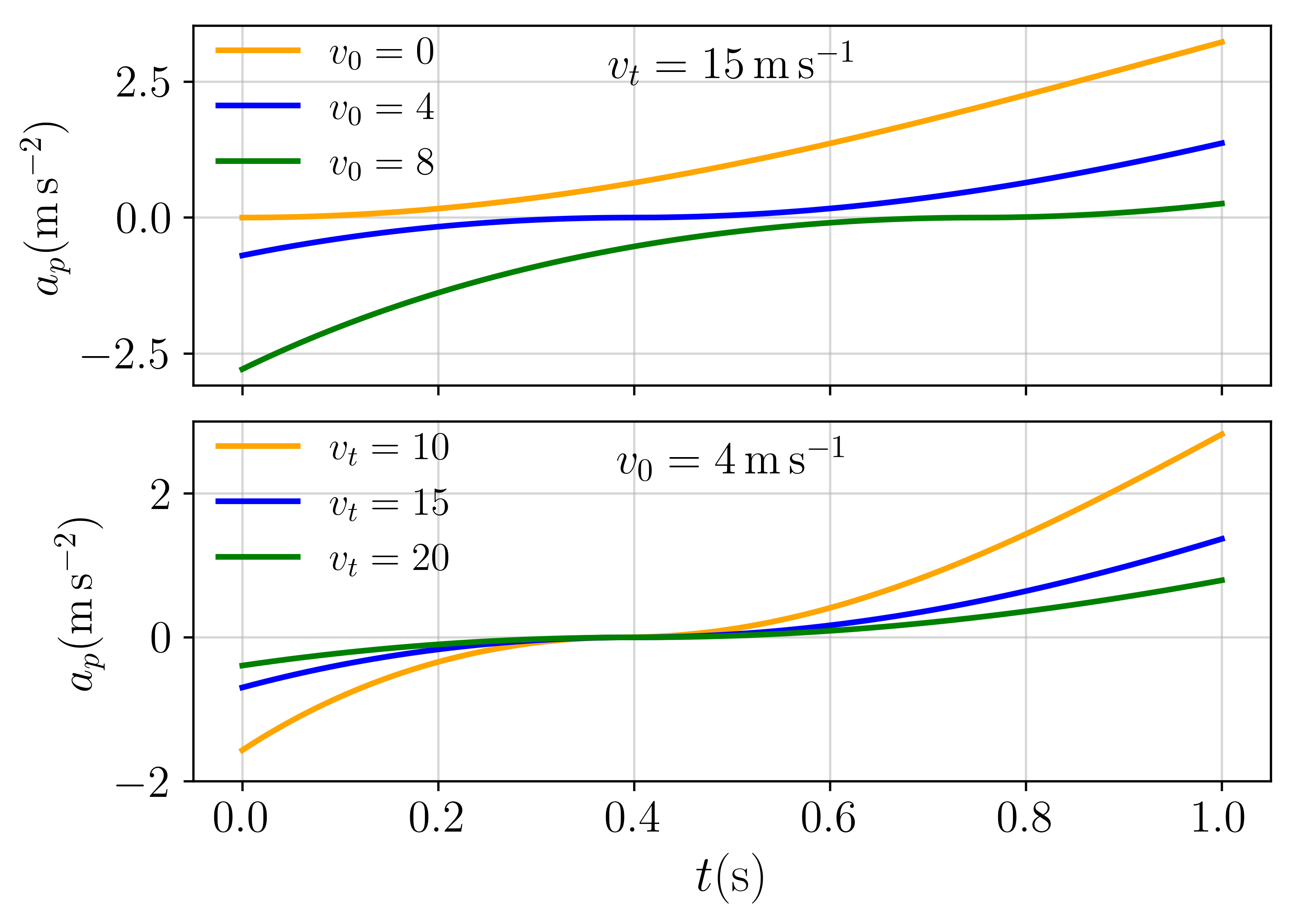}
    \caption{Proper acceleration \(a_p(t)\) in the turbulent flow regime for varying initial velocities \(v_0\) (top plot) and terminal velocities \(v_t\) (bottom plot). In the top plot, the terminal velocity is fixed at \(v_t = 15 \, \text{m} \, \text{s}^{-1}\), with initial velocities \(v_0 = 0 \, \text{m} \, \text{s}^{-1} \) (orange line), \(v_0 = 4 \, \text{m} \, \text{s}^{-1} \) (blue line), and \(v_0 = 8 \, \text{m} \, \text{s}^{-1}\) (green line). In the bottom plot, the initial velocity is fixed at \(v_0 = 4 \, \text{m}/\text{s}\), with terminal velocities \(v_t = 10 \, \text{m}/\text{s}\) (orange line), \(v_t = 15 \, \text{m} \, \text{s}^{-1}\) (blue line), and \(v_t = 20 \, \text{m} \, \text{s}^{-1} \) (green line).}
    \label{fig:ap_with_v0_vt}
\end{figure}

The top plot of Fig.~\ref{fig:ap_with_v0_vt} demonstrates the impact of different initial conditions on the proper acceleration in the turbulent drag model. It presents proper acceleration curves for varying initial velocities (\(v_{0} = 0 \, \text{m} \, \text{s}^{-1} \), orange line; \(v_{0} = 4 \, \text{m} \, \text{s}^{-1}\), blue line; and \(v_{0} = 8 \, \text{m} \, \text{s}^{-1}\)) while keeping the terminal velocity fixed at \(v_{t} = 15 \, \text{m} \, \text{s}^{-1}\).

Varying \(v_0\) results in a horizontal shift of the \(a_{p}(t)\) curve without changing its overall shape. This horizontal translation corresponds to a temporal shift in the acceleration profile, affecting the timing of key events, such as reaching the apex or approaching terminal velocity. However, the magnitude of the acceleration remains consistent across different initial velocity values.

The bottom plot of Fig.~\ref{fig:ap_with_v0_vt} demonstrates the effect of varying terminal velocities (\(v_{t} = 10 \, \text{m} \, \text{s}^{-1}\), orange line; \(v_{t} = 15 \, \text{m} \, \text{s}^{-1} \), blue line; and \(v_{t} = 20 \, \text{m} \, \text{s}^{-1} \)) while keeping the initial velocity fixed at \(v_{0} = 4 \, \text{m} \, \text{s}^{-1} \). As \(v_t\) increases, both the amplitude and the curvature of the \(a_p\) curve change, highlighting how terminal velocity influences the object's deceleration and eventual stabilization. 

Notably, while the curve’s shape varies, the time to reach the apex is only marginally affected, with a small shift of just 0.015 seconds across the different \(v_{t}\) values considered. This behavior can be understood by analyzing the Taylor series expansion of \( t_{\text{apex}} \) given by Eq.~\eqref{eq:t_apex_turbulent}:
\begin{equation}\label{eq:taylor_apex}
t_{\text{apex}} = \frac{v_0}{g} \left[ 1 - \frac{1}{3} \left( \frac{v_0}{v_t} \right)^{2} + \mathcal{O}\left(\frac{v_0}{v_t}\right)^{4} \right].
\end{equation}
The leading-order term represents the apex time when only gravity acts on the smartphone, independent of drag effects. The next term, proportional to \( \frac{1}{3}(v_0 / v_t)^2 \), quantifies the small correction due to the influence of terminal velocity, becoming significant only when \(v_0\) approaches \(v_t\).

\section{Data Acquisition}\label{sec:Data_Acquisition}

In this study, we used an Asus ZenFone Max Pro (M2)~\cite{gsmarena_zenfone_m2} smartphone to collect acceleration data during its fall. The accelerometer data were accessed and recorded using the phyphox application~\cite{Staacks_2018}, a free and open-source tool that provides real-time access to various sensor readings. This app allows for the simple export of collected data in multiple formats, making it suitable for detailed analysis.

\subsection{Experimental Setup}\label{sec:Experimental_Setup}

The experiments were designed with simplicity and replicability in mind, requiring minimal equipment. The smartphone was manually dropped or thrown vertically upward above a soft landing surface. In both cases, care was taken to keep the screen parallel to the Earth's surface and minimize unintended rotations, ensuring the phone's \( z_{p} \)-axis remained aligned with the Earth's \( z \)-axis.

The two types of experiments were:
\begin{itemize}
    \item Fall: In this experiment, the smartphone was released from rest with no significant initial velocity. A total of 22 trials were performed under these conditions.
    
    \item Up\(-\)Down: For this experiment, the smartphone was thrown vertically upward with an initial positive velocity. Data were collected throughout both the upward and downward motion phases. A total of 16 trials were conducted for this type of experiment.
\end{itemize}

Although the Up\(-\)Down experiment inherently includes a Fall phase, the two experiments were conducted separately to account for additional variability in the Up\(-\)Down setup. This setup is more prone to differences in initial velocity, as well as unintended inclinations and rotations during the launch. In contrast, the Fall setup, with negligible initial velocity and a more controlled release, potentially reduces these sources of variability.

\subsection{Accelerometer Precision and Variability}\label{sec:Accelerometer_Precision_and_Variability}

The Asus ZenFone Max Pro (M2) smartphone used in this experiment is equipped with a Bosch BMI160 accelerometer~\cite{bosch_bmi160}, a 6-axis motion sensor. For this experiment, only the linear acceleration data were used to analyze the smartphone's motion. The data were acquired at 200 Hz, the default setting in the phyphox application, providing adequate temporal resolution to capture the smooth evolution of the acceleration profile.

To evaluate the precision of the Bosch BMI160 accelerometer, two types of data collection were performed: Static and Free-Fall. In the Static method, the smartphone was placed stationary on a flat surface. In the Free-Fall method, the smartphone was dropped with its screen oriented perpendicular to the Earth's surface, \( \hat{z}_{p} \cdot \hat{z} = 0 \), to minimize drag. Note that this orientation differs from that used in the Fall and Up\(-\)Down experiments, where \( \hat{z}_{p} \cdot \hat{z} = 1 \). In both methods, the focus was on collecting acceleration data along the \( z_{p} \)-axis.

For the Static test, 1736 data points were collected over 8.75 seconds. The acceleration along the \(z_{p}\)-axis was measured as \(a_{p} = (9.71 \pm 0.02) \, \text{m} \, \text{s}^{-2}\), aligning closely with the expected gravitational acceleration. However, the observed deviation suggests the presence of an intrinsic offset in the accelerometer measurements, although a slight tilt relative to the vertical direction cannot be completely ruled out. This result is consistent with previous studies using smartphone accelerometers in similar setups~\cite{Monteiro21}.

For the Free-Fall tests, three trials were conducted, with the collected data summarized as follows:
\begin{itemize}
    \item Trial 1: 79 data points collected over 0.39 seconds, with an acceleration of \(a_{p} = (-0.25 \pm 0.02) \, \text{m} \, \text{s}^{-2}\).
    \item Trial 2: 95 data points recorded over 0.47 seconds, yielding an acceleration of \(a_{p} = (-0.24 \pm 0.02) \, \text{m} \, \text{s}^{-2}\).
    \item Trial 3: 104 data points obtained over 0.52 seconds, with an acceleration of \(a_{p} = (-0.24 \pm 0.02) \, \text{m} \, \text{s}^{-2}\).
\end{itemize}

Two important observations emerge from these results. The first is that the standard deviation is similar across all three trials and aligns with the value obtained in the Static test. This indicates that the accelerometer behaves similarly under both stationary and dynamic conditions. The second observation is that the acceleration values show close agreement across the trials, reinforcing the presence of an intrinsic bias in the measurements. These bias align with the zero-g offsets commonly observed in smartphone accelerometers, which can vary significantly between devices~\cite{Odenwald19}.

\begin{figure}[t]
    \centering
    \includegraphics[width=0.8\textwidth]{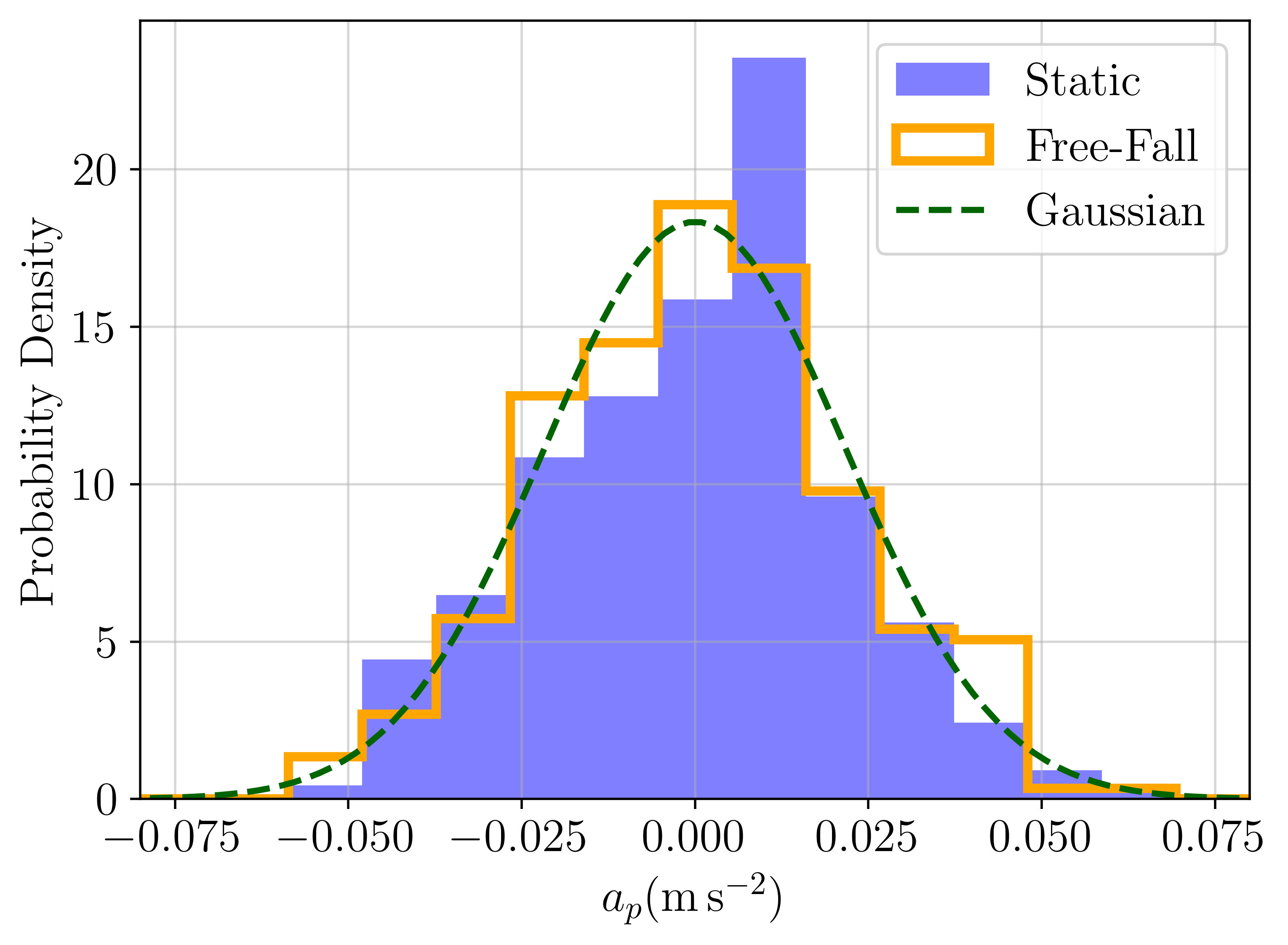}
    \caption{Probability density of proper acceleration \(a_p\) for both the Static (blue) and Free-Fall (orange) experiments, with their means subtracted. A Gaussian distribution with a standard deviation of \(0.02 \, \text{m} \, \text{s}^{-2}\) (dashed green) is also shown, reflecting the variability of the accelerometer across both experimental conditions.}
    \label{fig:accelerometer_random_uncertainty}
\end{figure}

Fig.~\ref{fig:accelerometer_random_uncertainty} shows the variability in the accelerometer readings from both the Static and Free-Fall experiments. For both setups, the mean acceleration was subtracted to focus only on the fluctuations. In the Free-Fall setup, the results from three trials were combined into a single dataset. Both distributions---Static (blue) and Free-Fall (orange)---show similar variabilities. A standard deviation of \(\sigma_{a_{p}} = 0.02 \, \text{m} \, \text{s}^{-2}\) was determined for the accelerometer, as indicated by the Gaussian curve (dashed green). This value will be used as the error in the accelerometer measurements for the following data analysis.

\subsection{Overview of Collected Data}\label{sec:Overview_of_Collected_Data}

Before analyzing all 38 datasets from the Fall (22 trials) and Up\(-\)Down (16 trials) experiments, we begin by focusing on a single Up\(-\)Down experiment. This initial analysis provides insight into both the upward and downward phases, since the Up\(-\)Down experiment inherently includes a Fall phase at its end.

\begin{figure}[t]
    \centering
    \includegraphics[width=0.8\textwidth]{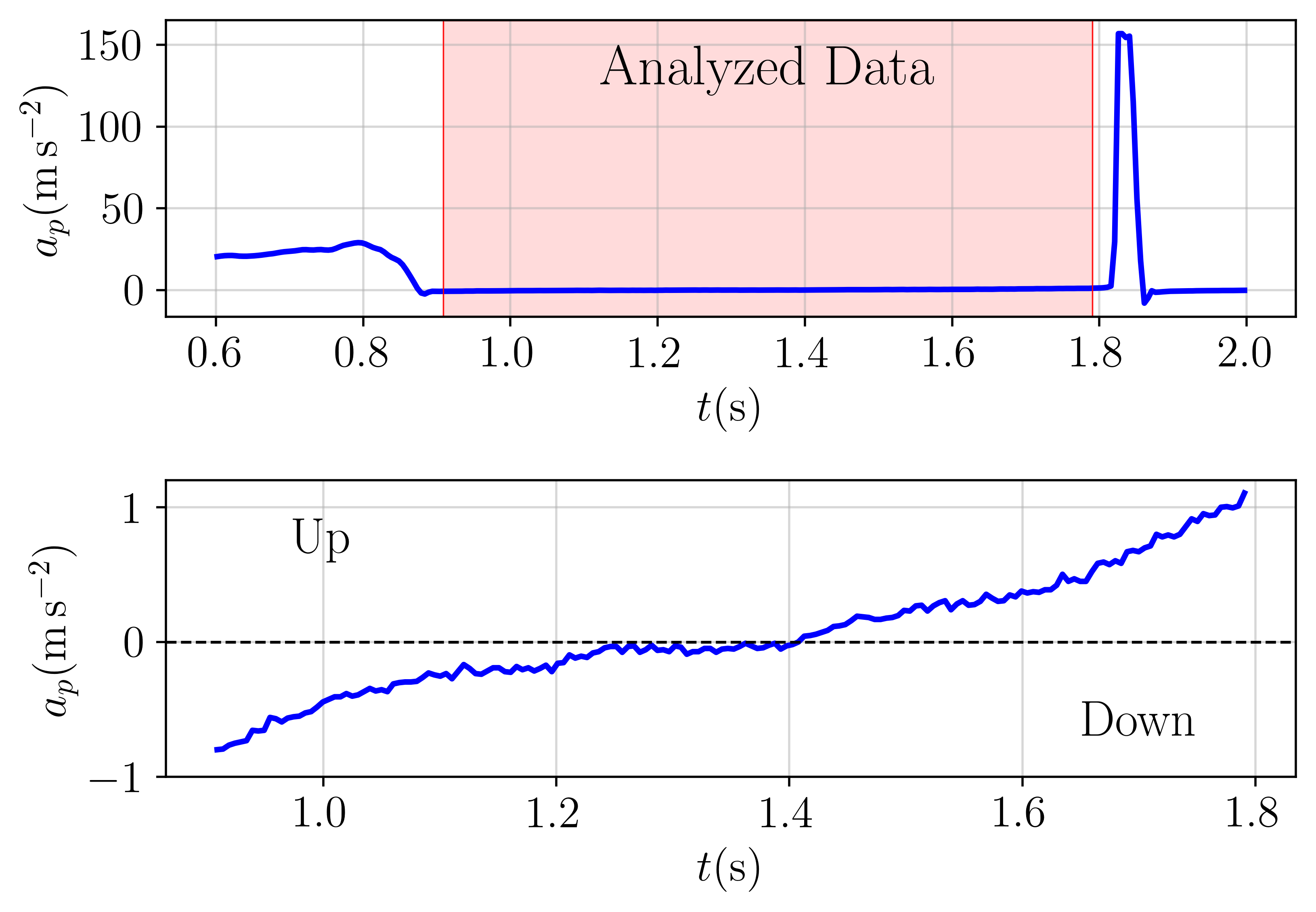}
    \caption{Raw acceleration data recorded by the smartphone's accelerometer (\(z_{p}\) component) during a trial of the Up\(-\)Down experiment. The upper plot shows the unfiltered data from just before the smartphone's release until its impact with the surface, with the shaded red region marking the interval selected for detailed analysis. The lower plot provides a zoomed-in view of this interval, with ``Up'' and ``Down'' annotations indicating the upward and downward phases as the smartphone transitions between them.}
    \label{fig:raw_data}
\end{figure}

Fig.~\ref{fig:raw_data} shows the acceleration data recorded by the smartphone's accelerometer during a typical Up\(-\)Down experiment. The upper plot presents the raw, unfiltered data along the \(z_{p}\)-axis, starting just before the smartphone is released and continuing until it impacts the surface. The upward launch is not instantaneous: it begins at 0.8 seconds, when the phone is released, and lasts until around 0.9 seconds, when the accelerometer readings stabilize. The impact is registered by a sharp peak in acceleration just after 1.8 seconds.

The red-highlighted section in the upper plot of Fig.~\ref{fig:raw_data} marks the period during which the smartphone is influenced only by gravity and air resistance. These data were selected through visual inspection, excluding any residual effects from the release or the impact. The analyzed region captures key characteristics of the motion, including the transition at the apex, where a change in the concavity of the acceleration curve is evident, as shown in the bottom plot of Fig.\ref{fig:raw_data}. 

We now expand our analysis to the full set of acceleration measurements from both experiments. By analyzing the 22 Fall and 16 Up\(-\)Down trials, we aim to capture the variability across them and gain a more comprehensive understanding of the smartphone's motion under varying experimental conditions.

\begin{figure}[t]
    \centering
    \includegraphics[width=0.9\textwidth]{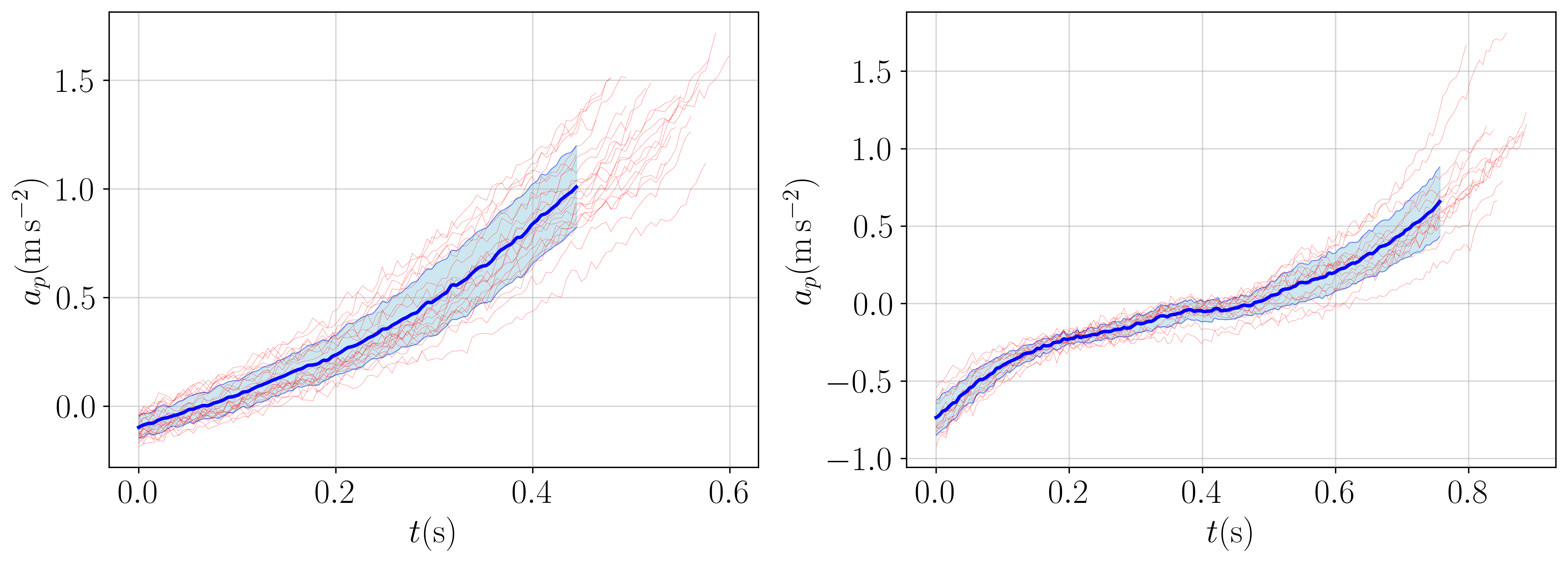}
    \caption{Proper acceleration (\(a_{p}\)) measured during the Fall (left) and Up\(-\)Down (right) experiments. Thin red lines represent individual measurements, while the blue line indicates the mean across the dataset. The shaded blue region corresponds to the standard deviation, offering a measure of variability across experiments.}
    \label{fig:fall_up_down_all_mean_std}
\end{figure}

In Fig.~\ref{fig:fall_up_down_all_mean_std}, the measured proper acceleration (\(a_{p}\)) is displayed for all experiments, with the Fall data in the left plot and the Up\(-\)Down data in the right plot. In both plots, thin red lines represent individual measurements, while the blue line shows the mean acceleration across the dataset. The shaded blue region around the blue line corresponds to the standard deviation, providing insight into the variability between measurements.

The variability observed across trials can be compared with the intrinsic variability of the accelerometer, determined as \( \sigma_{a_{p}} = 0.02 \, \text{m} \, \text{s}^{-2} \) (as defined in Section~\ref{sec:Accelerometer_Precision_and_Variability}). For the Fall experiment, the standard deviation across trials ranged from \( 0.05 \, \text{m} \, \text{s}^{-2} \) to \( 0.2 \, \text{m} \, \text{s}^{-2} \). Similarly, for the Up\(-\)Down experiment, the standard deviation showed values between \( 0.03 \, \text{m} \, \text{s}^{-2} \) and \( 0.2 \, \text{m} \, \text{s}^{-2} \). These values are notably larger than the intrinsic variability of the accelerometer, reflecting additional sources of variation inherent to the experimental setup, such as differences in initial velocity, tilt, and rotation across trials.

Interestingly, the variability observed across trials in both experiments is comparable, despite the Fall experiment being designed to minimize it. This suggests that initial conditions still introduce significant fluctuations, limiting the effectiveness of the Fall setup in reducing trial-to-trial variations.

\section{Data Analysis}\label{sec:Data_Analysis}

In this section, we analyze the acceleration data collected from the smartphone during the Fall and Up\(-\)Down experiments (see Section~\ref{sec:Experimental_Setup}) to validate the experimental setup and assess the applicability of the turbulent drag model.

\subsection{Methodology for Model Fitting}\label{sec:Methodology_for_Model_Fitting}

To fit the turbulent drag model to the experimental data, we employed the Weighted Least Squares (WLS) method, where the weights are given by the inverse square of the corresponding uncertainties~\cite{cowan98}. In this approach, the goal is to minimize the weighted sum of the squared residuals, denoted by \( \varepsilon_{i} \). The residuals are defined as the difference between the observed data points, \( a_{p}(t_{i}) \), and the corresponding values predicted by the model at each measured time, \( \hat{a}_{p}(t_{i}; \vec{\theta}) \):

\begin{equation}\label{eq:residuals}
\varepsilon_{i} = a_{p}(t_{i}) - \hat{a}_{p}(t_{i}; \vec{\theta}) .
\end{equation}
The free parameters of the model are encapsulated in \( \vec{\theta} \).

Our model, developed in Section~\ref{sec:model}, requires an additional free parameter to fully describe the data: the bias inherent in the accelerometer readings. This bias is evident in Figs.~\ref{fig:raw_data} and~\ref{fig:fall_up_down_all_mean_std}, and was also observed when assessing its variability in Section~\ref{sec:Accelerometer_Precision_and_Variability}. To account for this effect, we introduce an offset, \( \Delta a \), to our proper acceleration model (see Eq.~\ref{eq:accel_proper_vt}):

\begin{equation}\label{eq:model_offset}
\hat{a}_{p} = a_{p} + \Delta a .
\end{equation}

The model \( \hat{a}_{p} \), now fully defined, includes the gravitational acceleration \( g \), the initial velocity \( v_0 \), the terminal velocity \( v_t \), and the offset \( \Delta a \) to account for the accelerometer bias.

The proper acceleration given by Eqs.~\ref{eq:accel_proper_vt}, \ref{eq:v_turbulent_upward}, \ref{eq:t_apex_turbulent}, and~\ref{eq:v_turbulent_downward} shows that the gravitational acceleration \( g \) and the terminal velocity \( v_{t} \) always appear as ratios: \( g / v_{t} \) or \( g / v_{t}^{2} \). This reveals a degeneracy in the model, as changes in \( g \) can be compensated by corresponding changes in \( v_{t} \), making it difficult to independently determine both parameters from the data. In practical terms, this means that the fitting process may yield multiple combinations of \( g \) and \( v_{t} \) that produce similar results, leading to ambiguity in the interpretation of the model parameters.

To address this issue, we fixed the value of \( g \) at \( 9.8 \, \text{m} \, \text{s}^{-2} \), consistent with the gravitational acceleration near the Earth's surface. This choice eliminates the degeneracy, allowing us to focus on accurately estimating \( v_{t} \), the parameter directly related to air resistance affecting the smartphone's motion. As a result, our model now includes three free parameters:
\begin{equation}
\vec{\theta} = ( v_0, v_t, \Delta a ).
\end{equation}
These three parameters will be estimated in the subsequent analysis for both the Fall and Up\(-\)Down experiments.

\subsection{Fall Results}\label{sec:Fall_Results}

We begin by presenting the fitting results for the 22 individual trials from the Fall experiment. Each trial provided three fitted parameters: the initial velocity \( v_{0} \), the terminal velocity \( v_{t} \), and the offset \( \Delta a \). In total, 66 parameters were obtained across all trials.

\begin{figure}[t]
    \centering
    \includegraphics[width=0.8\textwidth]{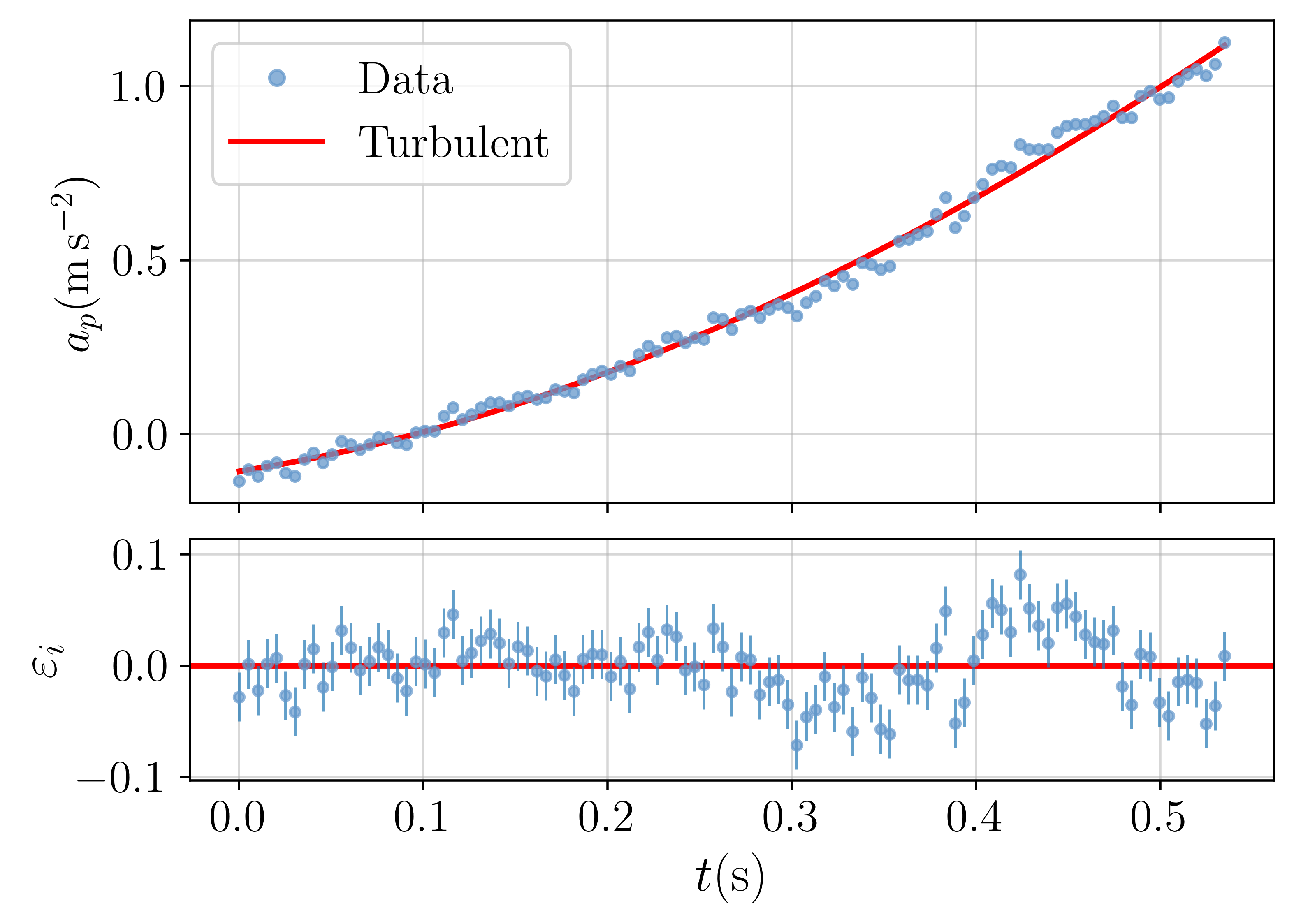}
    \caption{Fit of the turbulent drag model to a single dataset from the Fall experiment. The top plot displays the proper acceleration \( a_{p} \) as a function of time, showing experimental data (blue points) and the model fit (red line). The bottom plot illustrates the residuals \( \varepsilon_{i} \). Error bars indicate the uncertainty in the accelerometer measurements, \( \sigma_{a_{p}} = 0.02 \, \text{m} \, \text{s}^{-2} \) (error bars are omitted in the top plot for clarity). Further details of this fit are provided in the main text.}
    \label{fig:fall_fit_data_20}
\end{figure}

The fit of the turbulent drag model to a single dataset from the Fall experiment is shown in Fig.~\ref{fig:fall_fit_data_20}. This particular trial was selected because its reduced chi-square value, \( \chi^2/\text{dof} = 1.8 \), is close to the average value across all 22 trials, making it representative of the overall dataset. Although the terminal velocity, \( v_t = (17.3 \pm 0.3) \, \text{m} \, \text{s}^{-1} \), is the highest among all experiments, the model fits the data well, achieving a coefficient of determination \( R^2 = 0.994 \). The fitted initial velocity was \( v_0 = (-1.3 \pm 0.1) \, \text{m} \, \text{s}^{-1} \), indicating that the smartphone was already moving downward when data collection began. The offset was \( \Delta a = (-0.16 \pm 0.02) \, \text{m} \, \text{s}^{-2} \), reflecting the expected negative bias in the accelerometer readings.

The uncertainties in the fitted parameters, derived from the WLS method, exhibited significant variation across all 22 trials. For the initial velocity \(v_0\), the uncertainties ranged from \(0.04 \, \text{m} \, \text{s}^{-1}\) to \(0.2 \, \text{m} \, \text{s}^{-1}\), and for the terminal velocity \(v_t\), from \(0.1 \, \text{m} \, \text{s}^{-1}\) to \(0.3 \, \text{m} \, \text{s}^{-1}\). The offset in the proper acceleration, \(\Delta a\), which accounts for the accelerometer bias, had uncertainties between \(0.006 \, \text{m} \, \text{s}^{-2}\) and \(0.03 \, \text{m} \, \text{s}^{-2}\). These relatively large ranges of uncertainty result from the limitations in constraining the model in the Fall experiment. Since no data are available from the upward motion (and thus no information from the apex), the ability to refine the parameter estimates is inherently restricted. Notably, the three trials with positive initial velocities, although small (with the highest being just \(0.1 \, \text{m} \, \text{s}^{-1}\)), yielded significantly smaller parameter errors. This suggests that even slight upward motion provides additional information, improving the model's constraints.

The results of the 22 trials, when considered collectively, are summarized in Table~\ref{tab:fall_results}. The values presented correspond to the mean and standard deviation across all experiments, rather than the uncertainties derived from the WLS method. The coefficient of determination, \(R^2\), shows consistently high values, with \(0.995 \pm 0.002\), indicating good agreement between the experimental data and the turbulent drag model. The reduced chi-square, \(\chi^2/\text{dof} = 1.8 \pm 0.5\), reflects an overall good fit, further supporting the robustness of the model despite some variability between trials.

\begin{table}[t]
\centering
\begin{tabular}{lccc}
\hline
Parameter & Fall & Up\(-\)Down \\
\hline
\( R^{2} \) & \( 0.995 \pm 0.002 \) & \( 0.985 \pm 0.005 \) \\
\( \chi^2/\text{dof} \) & \( 1.8 \pm 0.5 \) & \( 5 \pm 3 \) \\
\( v_0 \, (\text{m} \, \text{s}^{-1}) \) & \( -0.8 \pm 0.6 \) & \( 3.6 \pm 0.4 \) \\
\( v_t \, (\text{m} \, \text{s}^{-1}) \) & \( 15 \pm 1 \) & \( 14 \pm 1 \) \\
\( \Delta a \, (\text{m} \, \text{s}^{-2}) \) & \( -0.11 \pm 0.06 \) & \( -0.08 \pm 0.06 \) \\
\hline
\end{tabular}
\caption{Summary of the results for the Fall and Up\(-\)Down experiments. The parameters include the coefficient of determination (\( R^2 \)), the reduced chi-square (\( \chi^2/\text{dof} \)), initial velocity (\( v_0 \)), terminal velocity (\( v_t \)), and the offset (\( \Delta a \)). The second column presents the mean and standard deviation of the results for the Fall experiments, while the third column summarizes the Up-Down experiments.}
\label{tab:fall_results}
\end{table}

Regarding the fitted parameters, the initial velocity \(v_0 = (-0.8 \pm 0.6) \, \text{m} \, \text{s}^{-1}\) shows considerable variability. This spread may be partly attributed to the experimental setup but is more likely due to the process of selecting the starting point for data collection, which could introduce variability in the determination of the initial velocity. Notably, in all but three trials, the initial velocity was negative, indicating that the smartphone had already started moving downward when data collection began. The terminal velocity \(v_t = (15 \pm 1) \, \text{m} \, \text{s}^{-1}\) showed notable variability across trials, reflecting unexpected fluctuations in a setup designed to minimize such effects. Nevertheless, the reduced chi-square value confirms a good agreement between the data and the turbulent drag model.

The offset in the proper acceleration, \( \Delta a = (-0.11 \pm 0.06) \, \text{m} \, \text{s}^{-2} \), exhibited a consistent trend, with only two trials yielding slightly positive values close to zero. This pattern confirms that the accelerometer’s bias was generally shifted toward negative values. Such small, stable offsets align with the expected behavior of the sensor, as discussed in Section~\ref{sec:Accelerometer_Precision_and_Variability}.

\subsection{Up-Down Results}\label{sec:Up_Down_Results}

\begin{figure}[t]
    \centering
    \includegraphics[width=0.8\textwidth]{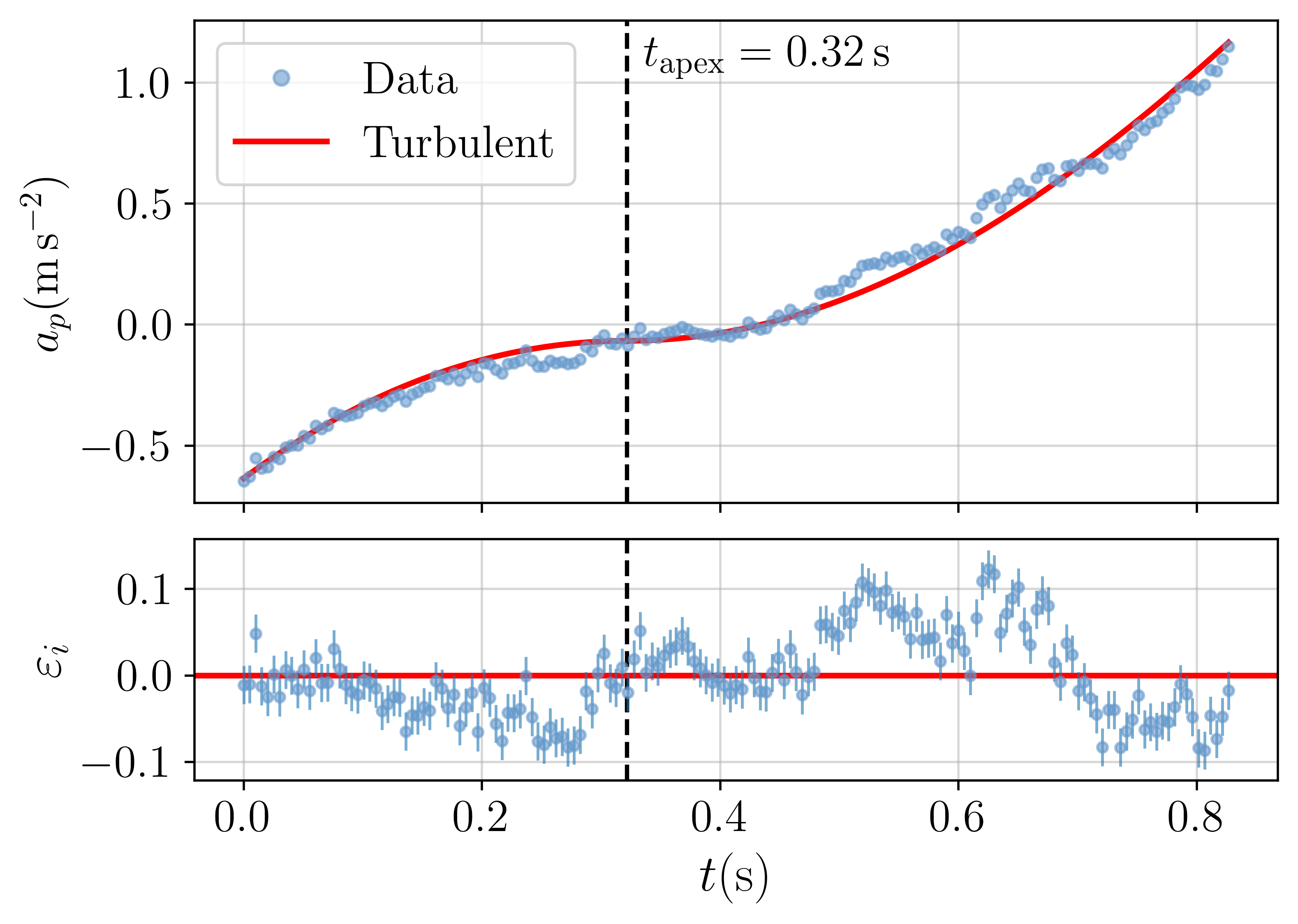}
    \caption{Fit of the turbulent drag model to a single dataset from the Up\(-\)Down experiment. The top plot shows the proper acceleration \( a_{p} \) as a function of time, with experimental data (blue points) and the model fit (red line). The dashed vertical line indicates the time \( t_{\text{apex}} \), marking the transition from upward to downward motion. The bottom plot presents the residuals \( \varepsilon_{i} \). The error bars represent the uncertainty in the accelerometer measurements, \( \sigma_{a_{p}} = 0.02 \, \text{m} \, \text{s}^{-2} \) (error bars are omitted in the top plot for clarity). The detailed results of this fit are discussed in the main text.}
    \label{fig:up_and_down_fit_data_7}
\end{figure}

The fitting results of the turbulent drag model for a single dataset from the Up\(-\)Down experiment are shown in Fig.~\ref{fig:up_and_down_fit_data_7}. This trial was selected because its chi-square value, \( \chi^2/\text{dof} = 5 \), is close to the average across all trials. Compared to the Fall setup, the higher chi-square value reflects the greater susceptibility of the Up\(-\)Down experiment to tilt and rotations. This result indicates that these effects are not negligible in this context and should be considered when interpreting the data.

Nonetheless, the model aligns well with the experimental data, achieving a coefficient of determination \( R^2 = 0.987 \). The fitted initial velocity was \( v_0 = (3.21 \pm 0.02) \, \text{m} \, \text{s}^{-1} \), indicating upward motion at release. The terminal velocity was \( v_t = (13.34 \pm 0.03) \, \text{m} \, \text{s}^{-1} \), and the offset in the proper acceleration was \( \Delta a = (-0.067 \pm 0.004) \, \text{m} \, \text{s}^{-2} \).

The uncertainties in the fitted parameters, derived from the WLS method, exhibited both minimal variation and small absolute values across the 16 individual trials. For the initial velocity \( v_0 \), they ranged from \( 0.01 \, \text{m} \, \text{s}^{-1} \) to \( 0.02 \, \text{m} \, \text{s}^{-1} \), while for the terminal velocity \( v_t \), the range was \( 0.02 \, \text{m} \, \text{s}^{-1} \) to \( 0.05 \, \text{m} \, \text{s}^{-1} \). The offset in the proper acceleration \( \Delta a \) had variations between \( 0.003 \, \text{m} \, \text{s}^{-2} \) and \( 0.004 \, \text{m} \, \text{s}^{-2} \). These small variations and low magnitudes reflect a more precise parameter estimation compared to the Fall experiment, where uncertainties were notably larger.

The results of the 16 individual trials from the Up\(-\)Down experiment are summarized in the third column of Table~\ref{tab:fall_results}. The coefficient of determination was \( R^2 = 0.985 \pm 0.005 \), which, although slightly lower than that observed in the Fall experiment, still indicates a reasonably good model fit. The reduced chi-square, \( \chi^2/\text{dof} = 5 \pm 3 \), was considerably higher, consistent with the increased susceptibility of the Up\(-\)Down setup to tilt and rotations, effects not accounted for in the model.

Regarding the fitted parameters, the initial velocity averaged 
\( v_0 = (3.6 \pm 0.4) \, \text{m} \, \text{s}^{-1} \), reflecting the upward 
motion of the smartphone at release. The small variability in \( v_0 \) across 
trials suggests a consistent launch between experiments. In contrast, the terminal 
velocity, \( v_t = (14 \pm 1) \, \text{m} \, \text{s}^{-1} \), showed 
greater variability, reflecting its sensitivity to minor changes in experimental 
conditions. The offset in proper acceleration, \( \Delta a = (-0.08 \pm 0.06) \, \text{m} \, \text{s}^{-2} \), 
remained consistently negative across most trials, indicating a slight bias in 
the accelerometer readings.

\subsection{Drag Coefficient}\label{sec:Drag_Coefficient}

The drag coefficient \((C_D)\) is a dimensionless quantity that measures the resistance an object experiences when moving through a fluid, such as air. At terminal velocity, the drag force balances the object’s weight, allowing us to calculate the drag coefficient for turbulent flow using the following relation~\cite{pettersson19}:

\begin{equation}
C_{D} = \frac{2 m g}{\rho A v_t^{2}} ,
\end{equation}
where \(m = 0.170 \, \text{kg}\) is the mass of the smartphone~\cite{gsmarena_zenfone_m2}, \(g = 9.8 \, \text{m} \, \text{s}^{-2}\) is the gravitational acceleration, \(\rho = 1.184 \, \text{kg} \, \text{m}^{-3}\) is the air density~\cite{engineersedge}, and \(A = L H\) is the cross-sectional area of the smartphone, with \(L = 7.55 \, \text{cm}\) and \(H = 15.79 \, \text{cm}\)~\cite{gsmarena_zenfone_m2}. The parameter \(v_t\) represents the terminal velocity.

The drag coefficient was calculated for both the Fall and Up\(-\)Down experiments using the terminal velocity values measured in each setup. The results were:
\begin{itemize}
    \item Fall: \( 1.1 \pm 0.2 \).
    \item Up\(-\)Down: \( 1.2 \pm 0.2 \),
\end{itemize}
showing a small difference and reasonable consistency between the setups.

The drag coefficients obtained in this study range from 0.8 to 1.7, aligning with reported values for similar objects, typically around 1.14~\cite{Fail1957}. Despite the variability observed between trials, these results demonstrate that our simple experimental setup can yield reliable estimates, comparable to those from more sophisticated fluid dynamics studies.

\section{Discussions and Conclusions}\label{sec:Discussions_and_Conclusions}

One notable aspect of this study is the accessibility and simplicity of its experimental setup. By leveraging the smartphone's built-in accelerometer, this experiment enables the exploration of air drag through a straightforward procedure. The process involves manually dropping or launching the smartphone vertically, ensuring simple data collection. No specialized equipment is required, making the setup practical for various contexts, including classroom environments or independent testing. Additionally, the simplicity of the procedure shifts the focus towards data analysis, offering valuable experience in experimental methods and insights into important aspects of data interpretation.

While the simplicity of the setup is advantageous, it also introduces challenges, particularly in controlling the smartphone’s launch with precision. Variability in initial velocity, tilt, or unintended rotations can affect the acceleration profiles. However, these variations provide an opportunity to explore how experimental conditions influence the results, emphasizing the importance of repeated measurements and careful consideration of uncertainties during data interpretation.

A total of 22 Fall and 16 Up\(-\)Down experiments were conducted (see Section~\ref{sec:Experimental_Setup} for details), with the results summarized in Table~\ref{tab:fall_results}. The offset results remained consistent across all trials, confirming the negative bias previously identified in the accelerometer measurements (see Section~\ref{sec:Accelerometer_Precision_and_Variability}). As expected, the initial velocity varied according to the different launching conditions: negative to near-zero values in the Fall experiments, \( (-0.8 \pm 0.6) \, \text{m} \, \text{s}^{-1} \), and positive in the Up\(-\)Down experiments, \( (3.6 \pm 0.4) \, \text{m} \, \text{s}^{-1} \).

\begin{figure}[t]
    \centering
    \includegraphics[width=0.8\textwidth]{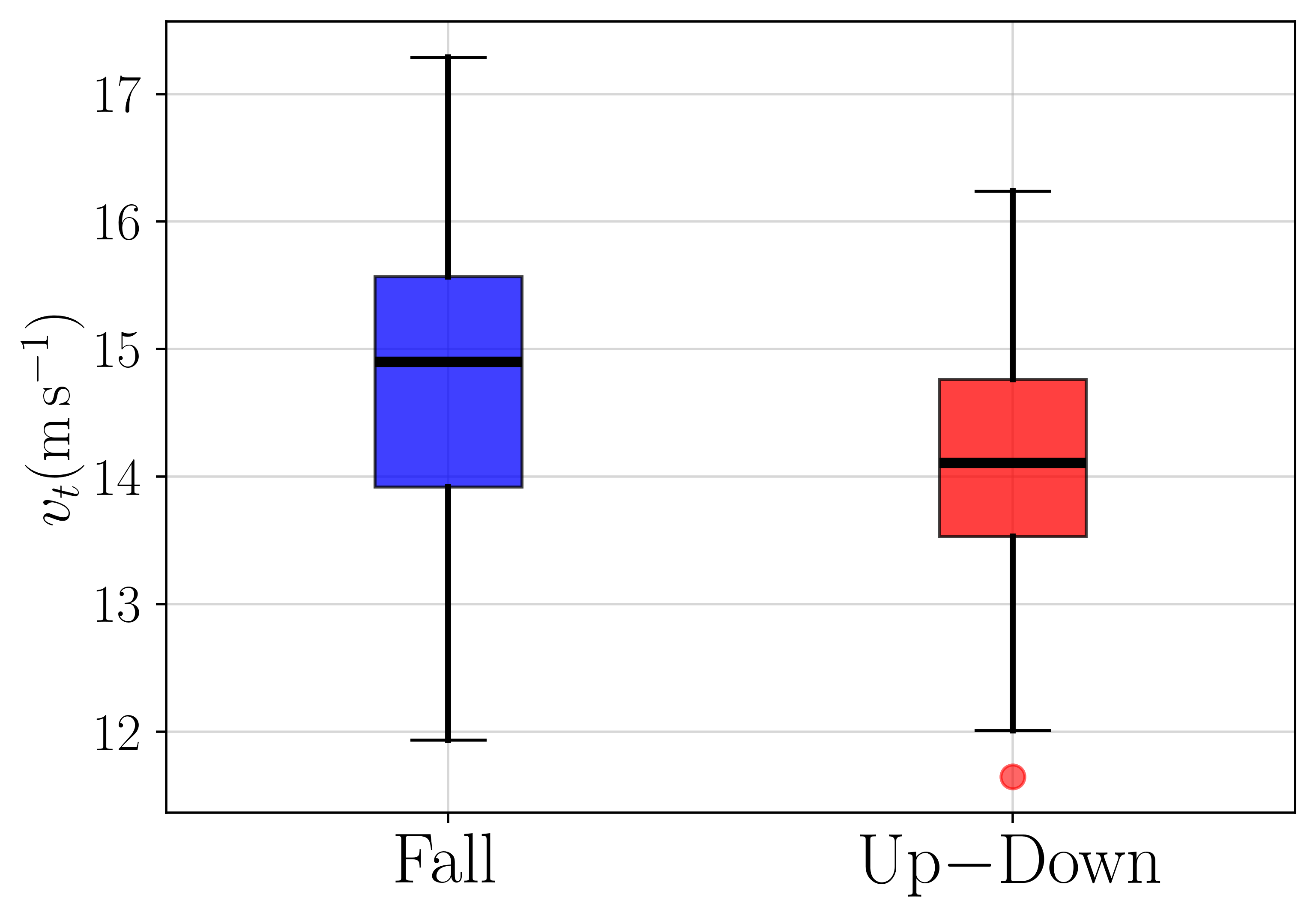}
    \caption{Comparison of terminal velocities \( v_{t} \) from the Fall (blue) and Up\(-\)Down (red) experiments. The boxes represent the interquartile range (IQR) of the individual trials, with the black horizontal line indicating the median value. The whiskers extend to the minimum and maximum observed values within 1.5 × IQR from the box edges, and individual points denote outliers (values beyond 1.5 × IQR).}
    \label{fig:v_t_box_plots}
\end{figure}

In order to provide a concise visual summary of the results, we present the box plot comparing the terminal velocity for both Fall and Up\(-\)Down experiments (see Fig.~\ref{fig:v_t_box_plots}). This box plot highlights the distribution of values for the individual trials, showing the interquartile ranges and median values, while also capturing the variability across experiments. The analysis of the terminal velocity showed consistent results across individual trials, with values around \(14 \, \text{m} \, \text{s}^{-1}\) (\(50 \, \text{km/h}\)) for both experimental setups. 

Despite the variability introduced by manual drops and the uncontrolled nature of the experiments, the turbulent drag model accurately described the observed motion in the Fall setup, as indicated by the reduced chi-square value close to 1. However, for the Up\(-\)Down setup, the significantly higher chi-square value highlights that tilt and rotations are considerable factors in this case. This distinction demonstrates the robustness of the approach in capturing the dynamics of air resistance while acknowledging the limitations imposed by experimental conditions.

\section*{Acknowledgements}

The author expresses sincere gratitude to AI tools, including ChatGPT, Claude, and Gemini, for their valuable assistance in refining research ideas, improving the structure of this manuscript, and contributing to the development of the Python code used in the analysis. Special thanks also go to the resilient smartphone, which endured countless falls for the sake of science. Though its journey ended in an unfortunate demise, its contribution will not be forgotten. The author is particularly grateful to one of the anonymous reviewers for their exceptionally detailed feedback, which greatly contributed to improving the quality of this work.


\begin{thebibliography}{6}

\bibitem{pagonis97} V. Pagonis, D. Guerra, S. Chauduri, B. Hornbecker, and N. Smiths, The Physics Teacher \textbf{35}, 364-368 (1997).

\bibitem{mohazzabi2011} P. Mohazzabi, The Physics Teacher \textbf{49}, 89-90 (2011).

\bibitem{mohazzabi2018} P. Mohazzabi, The Physics Teacher \textbf{56}, 168-169 (2018).

\bibitem{kundu2010fluid} P.K. Kundu and I.M. Cohen, \emph{Fluid Mechanics} (Academic Press, California, 2010), 2 ed.

\bibitem{anderson2010}  J.D. Anderson, \emph{Fundamentals of Aerodynamics} (McGraw-Hill Education, New York, 2010), 6 ed.

\bibitem{parker77} G. W. Parker, American Journal of Physics \textbf{45}, 606-610 (1977).

\bibitem{pettersson19} S. P. Fors and C. Nord, American Journal of Physics \textbf{87}, 714-719 (2019).

\bibitem{Wijaya_2019} P. A. Wijaya, U. Fauzi, and F. D. E. Latief, Physics Education, \textbf{54}, 055009 (2019).

\bibitem{vogt12} P. Vogt and J. Kuhn, The Physics Teacher \textbf{50}, 182-183 (2012).

\bibitem{Eager_2016} D. Eager, A. Pendrill, and N. Reistad, European Journal of Physics \textbf{37}, 065008 (2016).

\bibitem{gsmarena_zenfone_m2} GSMArena, \href{https://www.gsmarena.com/asus_zenfone_max_pro_(m2)_zb631kl-9410.php}{Asus ZenFone Max Pro (M2) ZB631KL Specifications}, accessed: 2024-10-02.   

\bibitem{Staacks_2018} S. Staacks, S. Hutz, H. Heinke, and C. Stampfer, Physics Education \textbf{53}, 045009 (2018).

\bibitem{bosch_bmi160} Bosch Sensortec, \href{https://www.bosch-sensortec.com/products/motion-sensors/imus/bmi160/}{BMI160 - Low power inertial measurement unit}, accessed: 2024-10-02.

\bibitem{Monteiro21} M. Monteiro, C. Stari, C. Cabeza, and A. C. Marti, American Journal of Physics \textbf{89}, 477-481 (2021).

\bibitem{Odenwald19} S. F. Odenwald and C. M. Bailey, IEEE Access \textbf{7}, 148131-148141 (2019).

\bibitem{engineersedge} Engineers Edge, \href{https://www.engineersedge.com/physics/viscosity_of_air_dynamic_and_kinematic_14483.htm}{Viscosity of Air, Dynamic and Kinematic}, accessed: 2024-10-02.

\bibitem{cowan98} G. Cowan, \emph{Statistical Data Analysis} (Oxford University Press, New York, 1998), 1 ed.

\bibitem{Fail1957} R. Fail, J. A. Lawford, and R. C. W. Eyre, \href{https://reports.aerade.cranfield.ac.uk/handle/1826.2/3689}{UK Ministry of Aviation Technical report} (1957). 

\end{thebibliography}
\end{document}